\documentstyle[aps,twocolumn,epsfig]{revtex}
\begin{document}
\twocolumn[\hsize\textwidth\columnwidth\hsize\csname
@twocolumnfalse\endcsname \draft

\title{Effects of field modulation on Aharonov-Bohm cages in a
       two-dimensional bipartite periodic lattice}
\author{Gi-Yeong Oh$^{*}$}
\address{Department of Basic Science, Hankyong National University,}
\address{Kyonggi-do 456-749, Korea}
%\date{submitted \today}
\maketitle

\begin{abstract}
We study the effects of field modulation on the energy spectrum of an
electron in a two-dimensional bipartite periodic lattice subject to a
magnetic field. Dependence of the energy spectrum on both the period
and the strength of field modulation is discussed in detail. Our main
finding is that introducing field modulation drastically changes the
energy spectrum and the localization properties of the system appearing
in the absence of field modulation; the degeneracies induced by a
uniform magnetic field are broken and the resultant energy spectrum
shows a dispersive band structure, indicating that most of
Aharonov-Bohm cages become unbounded. The effects of field modulation
on the superconducting transition temperature and the critical current
in a wire network with the same geometry are also discussed.
\end{abstract}

]

\section{Introduction}

The physics of magnetically induced frustration in various
two-dimensional (2D) structures including square,$^{1-3}$
rectangular,$^{4,5}$ triangular,$^{6}$ honeycomb,$^{7,8}$
aperiodic,$^{9}$ quasiperiodic,$^{10,11}$ fractal,$^{12}$ and even
random$^{13}$ geometries has attracted much interest in
condensed-matter physics for several decades. Recently, Vidal and
co-workers$^{14}$ presented a new localization mechanism induced by a
uniform magnetic field for noninteracting electrons in a 2D bipartite
periodic hexagonal structure (the so-called $T_{3}$ geometry), where
the unit cell contains three sites, one sixfold coordinated (called the
`hub' site) and two threefold coordinated (called the `rim' sites).
Within the tight-binding (TB) approximation, they showed that, due to
fully destructive quantum interference, the eigenstates at half a
magnetic flux quantum per elementary rhombus (briefly, half a flux) are
extremely localized and bounded in Aharonov-Bohm (AB) cages. Very
recently, Abilio and co-workers$^{15}$ performed transport measurements
on a superconducting wire network with the $T_{3}$ geometry and
confirmed the field-induced localization effect by observing a
depression of the superconducting transition temperature $T_{c}$ and
the critical current $J_{c}$ at half a flux.

Since the calculations in Ref.~14 were performed under the basic
assumptions of (i) the perfectness and the infinite size of the $T_{3}$
geometry and (ii) the spatial uniformity of the magnetic field, it may
be very interesting to address a question of what happens under the
situations beyond the two assumptions. Of course, effects beyond the
assumption (i) were already discussed by the authors of Ref.~14; the
randomness (such as random modulations of hopping terms and
fluctuations in the tiling areas and in the transmission matrix along
the edges), which is inevitable in real systems, was expected to alter
or even destroy the phase matching essential for the field-induced
localization effect. In addition, the authors of Ref.~15 argued that
the incomplete suppression of the experimentally observed $J_{c}$ at
half a flux might be attributed to the network's finite size. However,
effects beyond the assumption (ii) are not examined yet and remain an
open question. Thus, in this paper, we would like to address the
question and clarify the effects of a nonuniform magnetic field on the
energy spectrum and the transport properties of the $T_{3}$ geometry at
rational fluxes, especially at half a flux. As a concrete and simple
example, we consider a magnetic field with a periodic modulation and
investigate the effect of field modulation on the stability of AB cages
and on the characteristics of $T_{c}$ and $J_{c}$.

The contents of this paper are organized as follows. An eigenvalue
equation taking into account field modulation is derived in Sec.~II.
Analytic and numerical results for the effects of field modulation on
the energy spectrum at rational fluxes and the localization properties
of the eigenstates at half a flux are presented in Sec.~III. The
effects of field modulation on $T_{c}$ and $J_{c}$ of a superconducting
wire network are also discussed in this section. Finally, Sec.~IV is
devoted to a summary.

\section{Field Modulation and the eigenvalue equation}

We consider an electron in a 2D rhombus tiling under a spatially
modulated magnetic field as
\begin{equation}
\vec{B}=\left[B_{0}+B_{mod}(x)\right]\hat{z},              \label{eq:1}
\end{equation}
where $B_{0}~(B_{mod})$ denotes the uniform (modulated) part of the
applied magnetic field. Among possible types of modulated fields, we
pay attention to the one-dimensional (1D) sine-modulated field as
\begin{equation}
  B_{mod}(x)=B_{1}\sin\left(\frac{2\pi x}{T_{x}}\right),   \label{eq:2}
\end{equation}
where $T_{x}$ is the period of the modulation along the $x$ direction.
Under the Landau gauge, the vector potential is given by
\begin{equation}
\vec{A}=\left(0,B_{0}x-\frac{B_{1}T_{x}}{2\pi}
        \cos{\left(\frac{2\pi x}{T_{x}}\right)},0\right).  \label{eq:3}
\end{equation}

The TB equation that describes an electron on a 2D lattice subject to a
magnetic field reads
\begin{equation}
 E\psi_{i}=\sum_{j}t_{ij}e^{i\gamma_{ij}}\psi_{j},         \label{eq:4}
\end{equation}
where $t_{ij}$ is the hopping integral between the nearest-neighboring
sites $i$ and $j$, and the phase factor $\gamma_{ij}$ is given by
\begin{equation}
 \gamma_{ij}=\frac{2\pi}{\phi_{0}}\int_{i}^{j}
 \vec{A}\cdot d\vec{l},                                    \label{eq:5}
\end{equation}
$\phi_{0}=hc/e$ being the magnetic flux quantum. For the sake of
simplicity, we set $t_{ij}=1$. The phase factor along the $x$ direction
in Fig.~1 is zero under the Landau gauge. Denoting the phase factor
along the upward direction on the line $1 (2, 3, 4)$ in Fig.~1 as
$\gamma_{1(2,3,4)}$, and using the wave function at a hub site as
$\psi(x,y)=e^{ik_{y}y}\psi(x)$, Eq.~(\ref{eq:4}) can be written as
\begin{eqnarray}
 (E^{2}-6)&&\psi(x)=\left[e^{i(\gamma_{4}+\kappa)}
      +e^{i(\gamma_{3}+\kappa)}\right]
      \psi\left(x+\frac{3a}{2}\right) \nonumber\\
   &&+\left[e^{2i(\gamma_{3}+\kappa)}+e^{2i(\gamma_{2}+\kappa)}
      \right]\psi\left(x\right) \nonumber\\
   &&+\left[e^{i(\gamma_{2}+\kappa)}+e^{i(\gamma_{1}+\kappa)}
      \right]\psi\left(x-\frac{3a}{2}\right)
      +{\rm H.c.},                                         \label{eq:6}
\end{eqnarray}
where $\kappa=\sqrt{3}k_{y}a/2$.

Denoting $\psi_{m}=\psi(x)$ at $x=3ma/2$, the phase factors can be
written as
\begin{eqnarray}
 \gamma_{1}&=&\frac{3\gamma}{2}\left(m-\frac{1}{2}\right)
   -\frac{\gamma}{2}-K\cos\left[\frac{3\pi a}{T_{x}}
   \left(m-\frac{5}{6}\right)\right], \nonumber\\
 \gamma_{2}&=&\frac{3\gamma}{2}\left(m-\frac{1}{2}\right)
   +\frac{\gamma}{2}-K\cos\left[\frac{3\pi a}{T_{x}}
   \left(m-\frac{1}{6}\right)\right], \nonumber\\
 \gamma_{3}&=&\frac{3\gamma}{2}\left(m+\frac{1}{2}\right)
   -\frac{\gamma}{2}-K\cos\left[\frac{3\pi a}{T_{x}}
   \left(m+\frac{1}{6}\right)\right], \nonumber\\
 \gamma_{4}&=&\frac{3\gamma}{2}\left(m+\frac{1}{2}\right)
   +\frac{\gamma}{2}-K\cos\left[\frac{3\pi a}{T_{x}}
   \left(m+\frac{5}{6}\right)\right],                      \label{eq:7}
\end{eqnarray}
where
\begin{equation}
 \gamma=2\pi f=\frac{2\pi\phi}{\phi_{0}},~
 K=\frac{\gamma\beta T_{x}^{2}}{\pi^{2}a^{2}}
      \sin{\left(\frac{\pi a}{2T_{x}}\right)},~
 \beta=\frac{B_{1}}{B_{0}},                               \label{eq:7a}
\end{equation}
$\phi(=\sqrt{3}B_{0}a^{2}/2)$ being the uniform background magnetic
flux through the elementary rhombus. Now let us define
\begin{equation}
 \mu_{m}^{\pm}=\frac{\gamma_{2}\pm\gamma_{1}}{2},~
 \mu_{m+1}^{\pm}=\frac{\gamma_{4}\pm\gamma_{3}}{2},~
 \nu_{m}^{\pm}=\gamma_{3}\pm\gamma_{2},                   \label{eq:7b}
\end{equation}
and
\begin{equation}
 \lambda=\frac{E^{2}-6)}{4}.                               \label{eq:8}
\end{equation}
Then, Eq.~(\ref{eq:6}) can be reduced to a 1D equation, which is a
generalized version of the eigenvalue equation derived in Ref.~14:
\begin{equation}
 \lambda\psi_{m}=A_{m+1}\psi_{m+1}+C_{m}\psi_{m}
                 +A_{m}\psi_{m-1},                         \label{eq:9}
\end{equation}
where
\begin{eqnarray}
 A_{m}&&=\cos(\mu_{m}^{-})\cos(\mu_{m}^{+}+\kappa), \nonumber \\
 C_{m}&&=\cos(\nu_{m}^{-})\cos(\nu_{m}^{+}+2\kappa),      \label{eq:10}
\end{eqnarray}
with
\begin{eqnarray}
 \mu_{m}^{-}&&=\frac{\gamma}{2}+K\sin\left(\frac{\pi a}{T_{x}}\right)
    \sin\left[\frac{3\pi a}{T_{x}}\left(m-\frac{1}{2}
    \right)\right],                                  \nonumber\\
 \mu_{m}^{+}&&=\frac{3\gamma}{2}\left(m-\frac{1}{2}\right)
    -K\cos\left(\frac{\pi a}{T_{x}}\right)\cos\left[
    \frac{3\pi a}{T_{x}}\left(m-\frac{1}{2}\right)\right],\nonumber\\
 \nu_{m}^{-}&&=\frac{\gamma}{2}+2K\sin\left(\frac{\pi a}{2T_{x}}
    \right)\sin\left(\frac{3\pi ma}{T_{x}}\right),   \nonumber\\
 \nu_{m}^{+}&&=3\gamma m-2K\cos\left(\frac{\pi a}{2T_{x}}\right)
    \cos\left(\frac{3\pi ma}{T_{x}} \right).              \label{eq:11}
\end{eqnarray}

A close inspection of Eqs.~(\ref{eq:9}) -- (\ref{eq:11}) for a reduced
rational flux $f=p/q$ with mutual primes $p$ and $q$ enables us to
write the Bloch condition along the $x$ direction as
\begin{equation}
 \psi_{m+N}=e^{iN\eta}\psi_{m} .                          \label{eq:12}
\end{equation}
Here, $\eta=3k_{x}a/2$ and $N$ is given by
\begin{equation}
 N=\left\{\begin{array}{lc}
          {\rm L.C.M.}(2T,2q),  &~~ q\neq3q', p\neq 2p' \\
          {\rm L.C.M.}(2T,q),   &~~ q\neq3q', p=2p' \\
          {\rm L.C.M.}(2T,2q'), &~~ q=3q'   , p\neq 2p' \\
          {\rm L.C.M.}(2T,q'),  &~~ q=3q'   , p=2p' \\
          \end{array}\right.                             \label{eq:13}
\end{equation}
where
\begin{equation}
 T=\left\{\begin{array}{lc}
      T_{x}, &~~T_{x}\neq 3T_{x}' \\
      T_{x}',&~~T_{x}=3T_{x}' ,
              \end{array}\right.                         \label{eq:14}
\end{equation}
$q'$, $p'$, and $T_{x}'$ being integers. Using Eqs.~(\ref{eq:9}) and
(\ref{eq:12}), we obtain the characteristic matrix as
\begin{equation}
  \left(\begin{array}{cccccc}
  C_{1} & A_{2} & 0 & 0 & \cdots & A_{1}{\rm e}^{-iN\eta} \\
  A_{2} & C_{2} & A_{3} & 0 & \cdots & 0 \\
  0 & A_{3} & C_{3} & A_{4} & \cdots & 0 \\
  \vdots & \vdots & \vdots & \vdots & \ddots & \vdots \\
  A_{1}{\rm e}^{iN\eta} & 0 & 0 & 0 & \cdots & C_{N}
              \end{array}\right) .                        \label{eq:15}
\end{equation}
Thus, by diagonalizing Eq.~(\ref{eq:15}), we can obtain $2N$ energy
eigenvalues $\{\pm(6+4\lambda_{i})^{1/2}; i=1,2,\dots,N\}$ for a given
$\vec{k}$ and the full energy spectrum by sweeping all the $\vec{k}$
points in the magnetic Brillouin zone (MBZ).

\section{Results and discussion}

\subsection{$E-f$ diagram}

When a periodic field modulation is introduced, a new length (i.e.,
$T_{x}$) is added into two characteristic lengths, the lattice constant
$a$ and the magnetic length $l=(\hbar/eB_{0})^{1/2}$. Thus, a subtle
interplay among them can change the energy band structure and the
localization properties of the system obtained in the absence of field
modulation. We plot in Fig.~2 the $E-f$ diagrams without and with field
modulation. In the calculations, we took $T_{x}=3$ and $q=101$ ($1\le
p\le100$), and swept $(20\times20)$ $\vec{k}$ points in the MBZ.
Comparing Fig.~2(b) with Fig.~2(a), we observe the following effects of
field modulation. First, the occurrence of overlapping between
neighboring subbands makes most of small gaps closed. This phenomenon
of subband broadening (or gap closing) clearly appears for large values
of $\beta$ and/or $f$. Second, introducing field modulation lowers the
symmetry of the energy spectrum. The translational symmetry [i.e.,
$E(f+1)=E(f)$] and the reflection invariance about $f=1/2$ [i.e.,
$E(f)=E(1-f)$] appearing in a uniform magnetic field are no longer held
in the presence of field modulation. Note, however, that the energy
spectrum still exhibits the symmetry with respect to $E=0$ and the
reflection symmetry about $f=0$. The former reflects the bipartite
geometry of the rhombus tiling and the latter implies that there is no
way for an electron to discern the direction of the magnetic field.

\subsection{Energy dispersion at half a flux}

Before proceeding further, it may be worth noting that there are two
types of localization effects in the $T_{3}$ geometry. One is the
topological localization appearing at $E=0$, which originates from the
local topology of the lattice.$^{16}$ Since this localization is well
understood and independent of the magnetic field, we omit further
discussion on this phenomenon. The other is the uniform field-induced
localization appearing at $E=\pm\sqrt{6}$ at half a flux.$^{14}$ We pay
attention to how this kind of localization is influenced by introducing
field modulation. Note also that, owing to the reflection symmetry
about $E=0$, we pay attention to the energy spectrum only with $E>0$ in
the discussion below.

\subsubsection{Case of $T_{x}=2$}

In this case, $N=4$ and direct diagonalization of the $(4\times4)$
Hamiltonian matrix yields the following energy dispersion:
\begin{equation}
 E(k_{x},k_{y})=\pm\sqrt{6\pm4\lambda_{\pm}},             \label{eq:16}
\end{equation}
where
\begin{eqnarray}
 \lambda_{\pm}=&&\{(A_{+}^{2}+A_{-}^{2}+A_{0}^{2}/2
    )^{2}\pm[A_{0}^{2}(A_{+}^{2}+A_{-}^{2}  \nonumber\\
    &&+A_{0}^{2}/4)^{2}+4A_{+}^{2}A_{-}^{2}\sin^{2}(2\eta)]^{1/2}
    \}^{1/2},                                             \label{eq:17}
\end{eqnarray}
with
\begin{eqnarray}
 A_{0}&&=-\sin\left(\frac{4\beta}{\pi}\right)\cos
           \left(2\kappa\right),                   \nonumber\\
 A_{\pm}&&=\pm\sin\left(\frac{2\beta}{\pi}\right)\cos
 \left(\kappa\mp\frac{\pi}{4}\right).                     \label{eq:18}
\end{eqnarray}
A close inspection of Eq.~(\ref{eq:16}) shows that the upper and lower
edges of the energy dispersion with $E>0$ are given by
\begin{equation}
 E_{u}=\left[6+4\sqrt{2D_{1}}\right]^{1/2},~~~
 E_{d}=\left[6-4\sqrt{2D_{1}}\right]^{1/2},               \label{eq:19}
\end{equation}
and the bandwidth is given by
\begin{equation}
 \Delta=4\sqrt{3}\left[1-\left(1-\frac{8D_{1}}{9}\right)^{1/2}
            \right]^{1/2},                               \label{eq:19a}
\end{equation}
where
\begin{equation}
 D_{1}=\sin^{2}\left(\frac{2\beta}{\pi}\right)\left[1+2\cos^{2}
        \left(\frac{2\beta}{\pi}\right)\right].           \label{eq:20}
\end{equation}
Equation~(\ref{eq:19a}) clearly shows that the bandwidth varies from
$0$ to $4\sqrt{3}$ with varying $\beta$. Note that this is a remarkable
effect of field modulation; the modulated field breaks the degeneracy
at $E=\sqrt{6}$, which appears in a uniform magnetic field and makes
the energy spectrum dispersive.

\subsubsection{Case of $T_{x}=3$}

By means of a similar method for the case above, we have the energy
dispersion
\begin{equation}
 E(k_{x},k_{y})=\left\{\begin{array}{c}
 \pm\left\{6\pm4\left[D_{2}f(k_{x},k_{y})\right]^{1/2}\right\}^{1/2} \\
 \pm\left\{6\pm4\left[D_{2}g(k_{x},k_{y})\right]^{1/2}\right\}^{1/2},
                    \end{array}\right.                    \label{eq:21}
\end{equation}
where
\begin{eqnarray}
 f(k_{x},k_{y})&&=\cos^{2}\eta\cos^{2}\kappa
   +\sin^{2}\eta\sin^{2}\kappa,\nonumber\\
 g(k_{x},k_{y})&&=\sin^{2}\eta\cos^{2}\kappa
           +\cos^{2}\eta\sin^{2}\kappa                    \label{eq:22}
\end{eqnarray}
and
\begin{equation}
 D_{2}=\sin^{2}\left(\frac{9\sqrt{3}\beta}{4\pi}\right).  \label{eq:23}
\end{equation}
An inspection of Eq.~(\ref{eq:21}) shows that the upper and lower edges
of the energy dispersion with $E>0$ are given by
\begin{equation}
 E_{u}=\left[6+4\sqrt{2D_{2}}\right]^{1/2},~~~
 E_{d}=\left[6-4\sqrt{2D_{2}}\right]^{1/2},               \label{eq:24}
\end{equation}
and the bandwidth is given by the same form as Eq.~(\ref{eq:19a}) with
replacing $D_{1}$ by $D_{2}$, which varies from $0$ to $4\sqrt{2}$ with
varying $\beta$. We plot in Fig.~3 the energy spectrum as a function of
$\beta$, where the vertical lines are results obtained by numerically
diagonalizing Eq.~(\ref{eq:15}) and the boundary curves are obtained
directly from Eq.~(\ref{eq:24}). The inset shows the $\beta$ dependence
of the energy spectrum up to $\beta=6$. The formation of a dispersive
energy spectrum instead of a highly degenerate point spectrum can be
clearly seen in the figure.

Another remarkable effect of field modulation can be found in the
density of states (DOS). We plot in Fig.~4 the DOS for several values
of $\beta$. The DOS for $\beta=0$ consists of a $\delta$-function peak
at $E=\sqrt{6}$, as elucidated in Ref.~14. However, when $\beta\neq0$,
the DOS at $E=\sqrt{6}$ vanishes and $E=\sqrt{6}$ becomes an edge that
connects the two subbands whose DOS's exhibit the well-known pagoda
shapes with a logarithmic singularity in the middle of the subbands.
The formation of a dispersive energy spectrum and the vanishing of the
DOS at $E=\sqrt{6}$ indicate that the modulated field makes most of AB
cages unbounded, which in turn implies that the transport properties of
the system at half a flux will exhibit quite different behaviors from
those under a uniform magnetic field.

\subsubsection{Case of $T_{x}\ge 4$}

As an example to illustrate the fact that the energy spectrum
sensitively depends not only on $\beta$ but also $T_{x}$, we plot in
Fig.~5 the energy spectrum for $T_{x}=4$ as a function of $\beta$. In
this case, $N=8$ and there are eight subbands with $E>0$, some of which
may overlap or may have finite gaps between them depending on $\beta$.
Figure~5 shows that the number of distinguishable subbands runs from 1
to 6 with increasing $\beta$ up to 1. We found that the occurrence of
subband splitting (or gap opening) is a generic feature of the energy
spectrum at half a flux for $T_{x}\ge4$. We also found that the subband
splitting occurs more easily when $T_{x}\neq3T_{x}'$ than when
$T_{x}=3T_{x}'$.

The $T_{x}$ dependence of the DOS also exhibits an interesting feature.
For illustrative purposes, we plot in Fig.~6 the DOS for $T_{x}=4$ for
several values of $\beta$. It can be seen that $E=\sqrt{6}$ locates
near an edge of a subband for small values of $\beta$, while it locates
in a subband for large values of $\beta$. Note, however, that even in
the latter case the DOS at $E=\sqrt{6}$ is still negligibly small
compared with the integrated DOS and will have a very little effect on
the transport properties of the system, if any.

\subsection{Energy dispersion at generic rational fluxes}

The $\beta$ dependence of the energy spectra for generic rationals $f$
exhibits a behavior similar to the case of $f=1/2$; the phenomena of
subband broadening and gap opening are generic features under the
modulated field. We show an example in Fig.~7, where the energy
spectrum with $E>0$ is plotted for $f=2/3$ and $T_{x}=4$. When
$\beta=0$, the energy spectrum consists of two subbands that touch each
other at $E=0$, resulting in a gapless single-band structure.$^{14}$
However, when the field modulation is turned on, a gap opens between
the two subbands. Besides, with increasing $\beta$, there occurs
further splitting of each subband into several sub-subbands as well as
the gap closing and reopening.

\subsection{Effect of $\beta$ on $T_{C}$ in a wire network}

Now we are in position to discuss the effects of field modulation on
$T_{c}$ and $J_{c}$ in a wire network. In the case of the $T_{3}$
geometry, $T_{c}(f)$ is directly related to the edge eigenvalue of
Eq.~(\ref{eq:9}) by$^{2,15}$
\begin{equation}
 1-\frac{T_{c}(f)}{T_{c}(0)}=C\left[\arccos
 \left(\frac{E_{e}(f)}{\sqrt{18}}\right)\right]^{2},      \label{eq:25}
\end{equation}
where $C$ is a constant that is proportional to the square of the
superconducting coherence length at zero temperature and $E_{e}(f)$ is
the edge eigenvalue at a rational flux $f$. For simplicity, we set
$C=1$.

We plot in Fig.~8 the phase boundaries for a system with $T_{x}=3$. In
the calculations, we took $q=101$ ($1\le p\le100$). It can be clearly
seen that the phase boundary for $\beta=0$ is symmetric about $f=1/2$
and distinct downward cusps occur at low order rationals $f=1/3, 2/3,
1/6, 5/6, 2/9$, and $7/9$. Besides, an upward cusp exists at $f=1/2$,
which reflects the field-induced localization properties of the
eigenstates at half a flux, as demonstrated in Ref.~15. However, when
the field modulation is introduced, the phase boundary exhibits several
distinctive features from the case of $\beta=0$ as follows. First, the
phase boundary becomes asymmetric with respect to $f=1/2$, which
reflects the breaking of the reflection invariance of the energy
spectrum about $f=1/2$. Second, the phase boundary for $f\ge1/3$
severely changes in a nontrivial way even for small values of $\beta$,
while the boundary for $f\le1/3$ is less influenced even by large
values of $\beta$. To put it concretely, of the cusps at strong
commensurate fields exhibited in the absence of field modulation, the
cusps at $f=1/3$ and $1/6$ are still clearly visible, while the cusp at
$f=5/6$ becomes invisible by introducing field modulation. Meanwhile,
the cusp at $f=2/3$ is visible (invisible) at $\beta=0, 0.6$
($\beta=0.3$). Third, with increasing $\beta$, the $T_{c}(f)$
depression up to $f\le4/5$ decreases independently of $T_{x}$. The
$\beta$ dependence of the $T_{c}$ depression at $f=1/2$ is shown in
Fig.~9, where a monotonic decrease of $1-T_{c}(f)/T_{c}(0)$ is clearly
seen. Meanwhile, the amount of the $T_{c}$ depression for $f\ge4/5$ is
inconsistent with the modulation strength. Fourth, the cusp at $f=1/2$
moves downward by introducing a field modulation, which is the most
remarkable effect of field modulation on $T_{c}$. This phenomenon
indicates that the localization properties of the system at $f=1/2$
becomes similar to those at other rational fluxes (such as
$f=1/3,1/2,1/6,5/6$) in which the eigenstates exhibit an extended
nature. Fifth, the value of $f$ where the maximal $T_{c}$ depression is
achieved lowers as $\beta$ increases.

Now, let us briefly discuss the effect of field modulation on the
critical current $J_{c}(f)$, which is closely related to the band
curvature, $\partial^{2}E_{e}(f)/\partial k^{2}$, near the band edge.
When $\beta=0$, due to the absence of dispersive states, $J_{c}(f=1/2)$
vanishes completely, as discussed in Ref.~15. However, when
$\beta\neq0$, the formation of a dispersive band structure makes
$\partial^{2}E_{e}(f)/\partial k^{2}\neq0$, and hence $J_{c}(f=1/2)$
will have a finite value, which is another remarkable effect of field
modulation.

\section{Summary}

We have studied the effects of field modulation on the energy spectrum
of an electron in a 2D bipartite periodic lattice subject to a magnetic
field and on the superconducting transition temperature in a wire
network with the same geometry. We have shown that the energy spectrum
sensitively depends on both the period and the strength of field
modulation. Our main finding is that the field-induced localization
properties of the lattice at half a flux are drastically changed by
introducing field modulation; the modulated field breaks the degeneracy
induced by a uniform magnetic field to make the energy spectrum
dispersive, where the number of distinguishable subbands sensitively
depends on both the period and the strength of field modulation. The
formation of a dispersive energy spectrum in turn has been shown to
crucially influence the superconducting transition temperature and the
critical current of the wire network. Before concluding this paper, we
would like to make a few remarks. First, though we treated only the
$T_{3}$ geometry in this paper, we expect that the field-induced
localization properties of the $T_{4}$ geometry$^{14}$ will undergo
similar effects of field modulation. Second, for the sake of simplicity
in the calculation, we dealt with a simple case where the period of
field modulation is commensurate with the lattice period. The energy
spectrum may exhibit more complicated band structures when the two
length scales are incommensurate with each other. Finally, note that
the magnetic flux treated in this paper is neither uniform nor random
but periodic. Thus, the effects of a random magnetic flux is still an
open question, and it would be interesting to study whether or not AB
cages remain bounded and/or how the localization properties of the
system are influenced by introducing a random flux. Our naive
expectation is that switching on a random flux at $f=1/2$ will induce
an energy band of localized states.

\acknowledgments{This work was financially supported by Hankyong
National University, Korea, through the program year of 1999.}

%--------------- References

%--------------- Figures

\begin{figure}[ht]
\centerline{\epsfig{figure=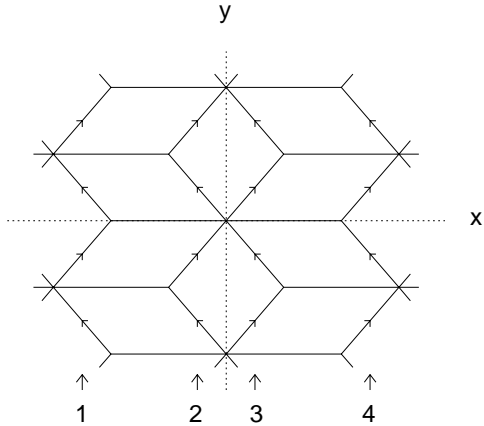,width=2.5in}}
\caption{ A portion of the 2D bipartite periodic tiling. \label{fig1}}
\end{figure}

\begin{figure}[ht]
\centerline{\epsfig{figure=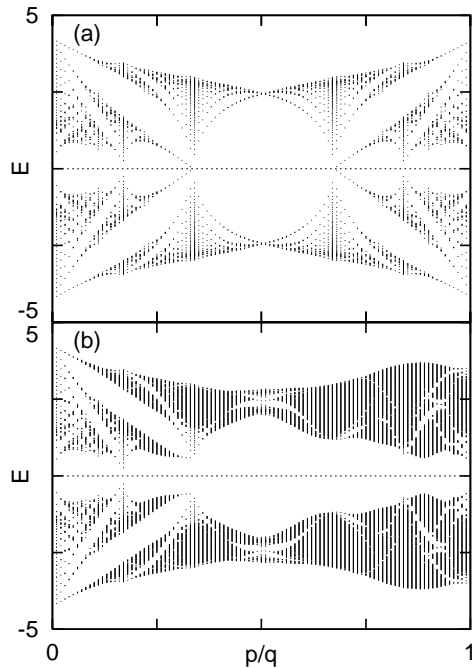,width=2.5in}}
\caption{Energy spectrum as a function of $f$ for $T_{x}=3$. (a)
         $\beta=0.0$, (b) $\beta=0.3$. \label{fig2}}
\end{figure}

\begin{figure}[ht]
\centerline{\epsfig{figure=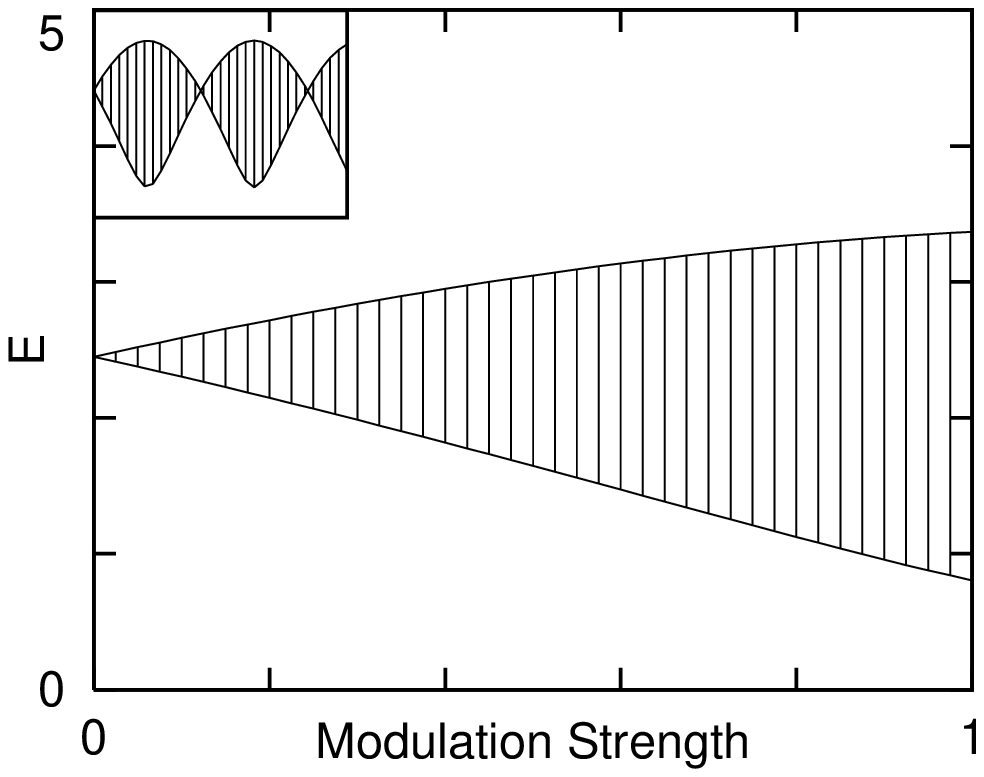,width=2.5in}}
\caption{Energy spectrum as a function of $\beta$ for $f=1/2$ and
         $T_{x}=3$. \label{fig3}}
\end{figure}

\begin{figure}[ht]
\centerline{\epsfig{figure=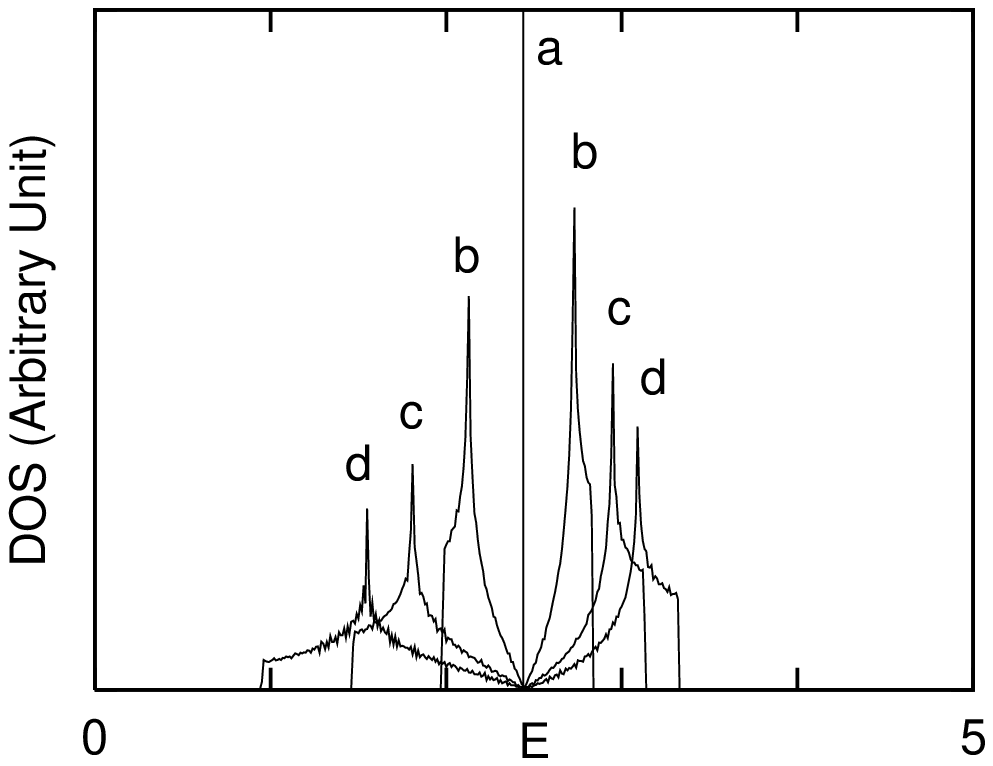,width=2.5in}}
\caption{Density of states for $f=1/2$ and $T_{x}=3$. (a) $\beta=0.0$,
        (b) $\beta=0.3$, (c) $\beta=0.6$, (d) $\beta=0.9$.\label{fig4}}
\end{figure}

\begin{figure}[ht]
\centerline{\epsfig{figure=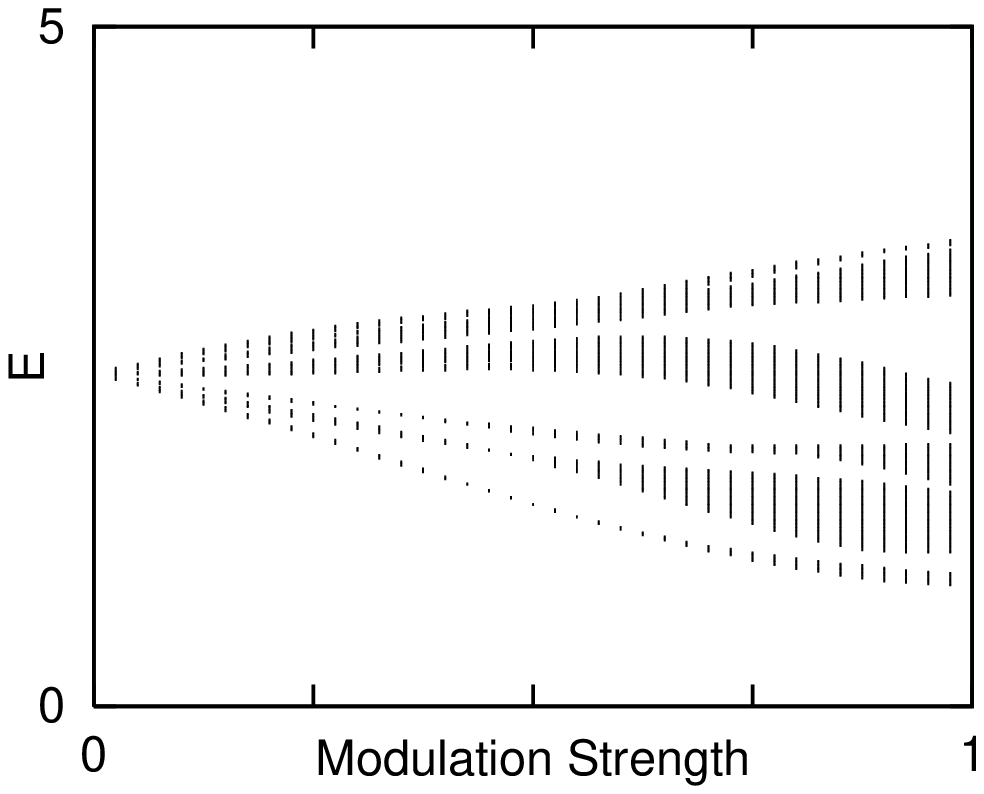,width=2.5in}}
\caption{Energy spectrum as a function of $\beta$ for $f=1/2$ and
         $T_{x}=4$. \label{fig5}}
\end{figure}

\begin{figure}[ht]
\centerline{\epsfig{figure=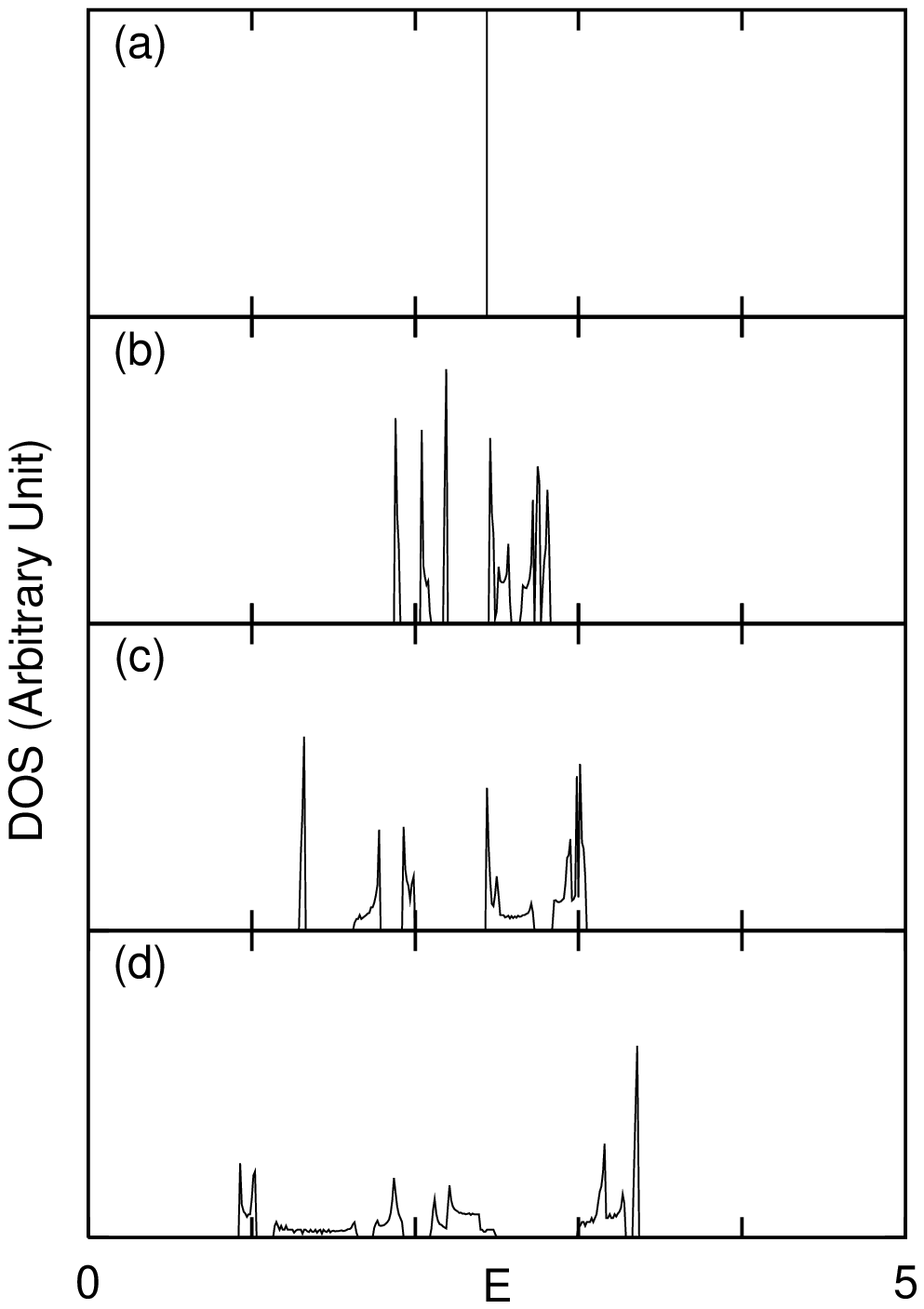,width=2.5in}}
\caption{Density of states for $f=1/2$ and $T_{x}=4$. (a) $\beta=0.0$,
        (b) $\beta=0.3$, (c) $\beta=0.6$, (d) $\beta=0.9$.\label{fig6}}
\end{figure}

\begin{figure}[ht]
\centerline{\epsfig{figure=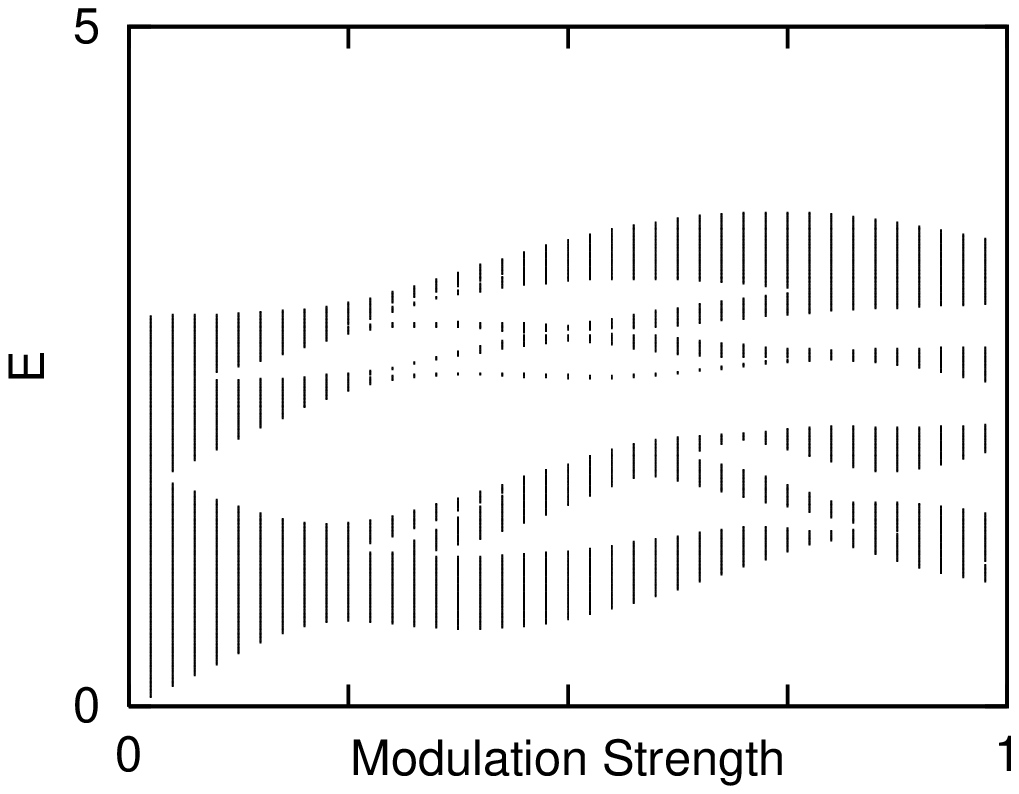,width=2.5in}}
\caption{Energy spectrum as a function of $\beta$ for $f=2/3$ and
         $T_{x}=4$. \label{fig7}}
\end{figure}

\newpage

\begin{figure}[ht]
\centerline{\epsfig{figure=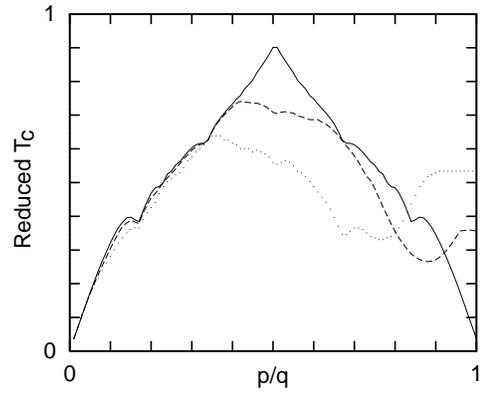,width=2.5in}}
\caption{$1-T_{c}(f)/T_{c}(0)$ as a function of $f$ for $T_{x}=3$.
         Continuous line: $\beta=0$, dashed line: $\beta=0.3$, dotted
         line: $\beta=0.6$. \label{fig8}}
\end{figure}

\begin{figure}[ht]
\centerline{\epsfig{figure=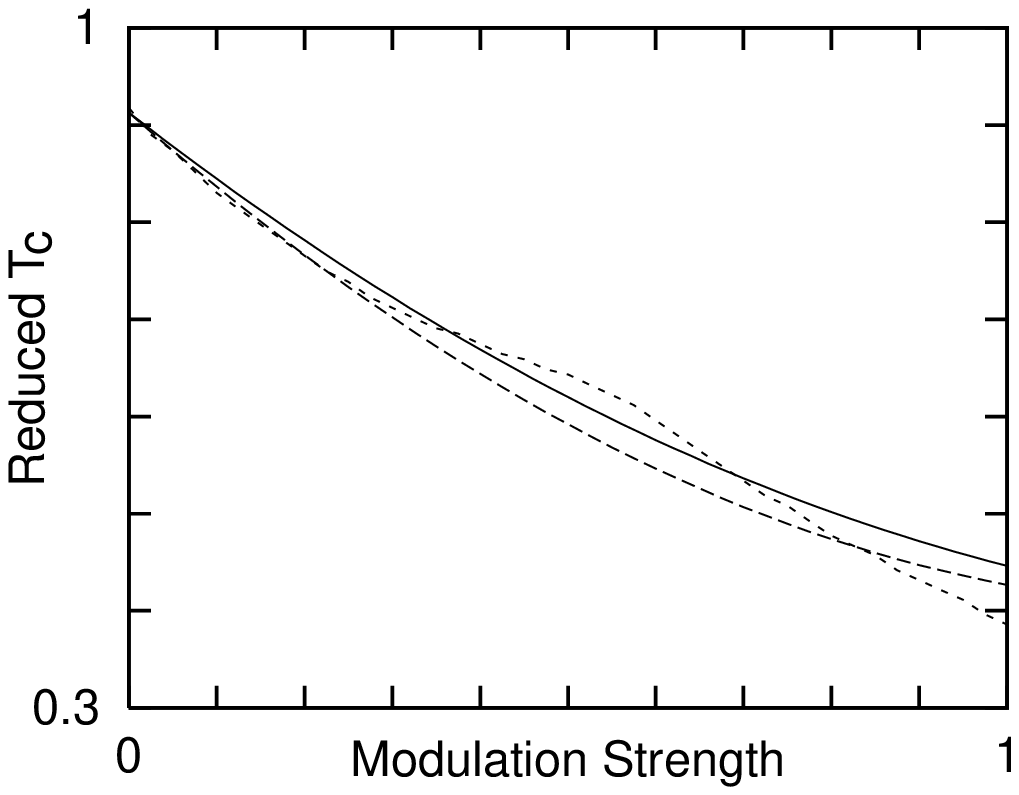,width=2.5in} }
\caption{$1-T_{c}(f)/T_{c}(0)$ as a function of $\beta$ for $f=1/2$.
         Continuous line: $T_{x}=2$, long-dashed line: $T_{x}=3$,
         short-dashed line: $T_{x}=4$. \label{fig9}}
\end{figure}


\begin{references}
\bibitem[*]{0} Electronic address: ogy@hnu.hankyong.ac.kr

%--- square E-spec theory
\bibitem{1}
 R. Peierls, Z. Phys. {\bf 80}, 763 (1933);
 P. G. Harper, Proc. Phys. Soc. London Sect. A {\bf 68}, 874 (1955);
 W. Kohn, Phys. Rev. {\bf 115}, 1460 (1959);
 M. Ya Azbel, Zh. Eksp. Teor. Fiz. {\bf 46}, 929 (1964) [Sov. Phys.
    JETP {\bf 19}, 634 (1964)];
 D. R. Hofstadter, Phys.  Rev. B {\bf 14}, 2239 (1976);
 G. H. Wannier, Phys. Status Solidi B {\bf 88}, 757 (1978).

%--- sc wire theory
\bibitem{2}
 P. G. de Gennes, C. R. Seances Aacd. Sci., Ser. B {\bf 292}, 9 (1981); {\bf
 292} 279 (1981);
 S. Alexander, Phys. Rev. B {\bf 27}, 1541 (1983);
 R. Rammal, T. C. Lubensky, and G. Toulose, {\it ibid}. {\bf 27},
    2820 (1983);
 S. N. Sun and J. P. Ralston, {\it ibid}. {\bf 43}, 5375 (1991).

%--- square wire experiment
\bibitem{3}
 B. Pannetier, J. Chaussy, R. Rammal, and J. C. Villegier, Phys. Rev.
    Lett. {\bf 53}, 1845 (1984);
 O. Buisson, M. Giroud, and B. Pannetier, Europhys. Lett. {\bf 12}, 727
 (1990);
 M. A. Itzler, R. Bojko, and P. M. Chaikin,  {\it ibid}. {\bf 20},
 639 (1992).

%--- rectangular E-spec theory
\bibitem{4}
 A. Barelli, J. Bellissard, and F. Claro, Phys. Rev. Lett. {\bf 83},
 5082 (1999).
%--- rectangular sc wire theory
\bibitem{5}
 C. R. Hu, Phys. Rev. B {\bf 35}, 5294 (1987);
 C. R. Hu and R. L. Chen, {\it ibid}. {\bf 37}, 7907 (1988).

%\vskip 3.65in
%\newpage

%--- triangular
\bibitem{6}
 D. Langbein, Phys. Rev. {\bf 180}, 633 (1969);
 F. H. Claro and G. H. Wannier, Phys. Rev. B {\bf 19}, 6068 (1979);
 D. J. Thouless, {\it ibid}. {\bf 28}, 4272 (1983);
 Y. Hasegawa, Y. Hatsugai, M. Kohmoto, and G. Montambaux, {\it ibid}.
    {\bf 41}, 9174 (1990);
 G. Y. Oh, J. Phys.: Condens. Matter {\bf 12}, 1539 (2000).

%--- honeycomb theory
\bibitem{7}
 T. Horiguchi, J. Math. Phys. {\bf 13}, 1411 (1972);
 R. Rammal, J. Phys. (France) {\bf 46}, 1345 (1985);
 V. A. Geyler and I. Yu Popov, Z. Phys. B {\bf 98}, 473 (1995).
%--- honeycomb experiment
\bibitem{8}
 B. Pannetier, J. Chaussy, and R. Rammal, J. Phys. (France) {\bf 44},
 L853 (1983).

%--- aperiodic wire theory and experiment
\bibitem{9}
 P. Santhanam, C. C. Chi, and W. W. Molzen, Phys. Rev. B {\bf 37}, 2360
 (1988);
 F. Nori and Q. Niu, {\it ibid}. {\bf  37}, 2364 (1988).

%--- quasiperiodic theory
\bibitem{10}
 M. Arai, T. Tokihiro, and T. Fujiwara, J. Phys. Soc. Jpn. {\bf 56},
 1642 (1987);
 T. Hatakeyama and H. Kamimura, {\it ibid}. {\bf 58}, 260
 (1989);
 Q. Niu and F. Nori, Phys. Rev. B {\bf 39}, 2134 (1989);
 M. A. Itzler, R. Bojko, and P. M. Chaikin, {\it ibid}. {\bf 47},
 14165 (1993);
 H. Schwabe, G. Kasner, and H. B\"{o}ttger, {\it ibid}. {\bf 56}, 8026
 (1997).

%--- quasiperiodic experiment
\bibitem{11}
 A. Behrooz, M. J. Burns, H. Deckman, D. Levine, B. Whitehead, and P.
 M. Chaikin, Phys. Rev. Lett. {\bf 57}, 368 (1986).

%--- fractal
\bibitem{12}
 R. Rammal and G. Toulouse, Phys. Rev. Lett. {\bf 49}, 1194 (1982);
 J. R. Banavar, L. Kadanoff, and A. M. M. Pruisken, Phys. Rev. B {\bf
 31}, 1388 (1985);
 X. R. Wang, {\it ibid}. {\bf 53}, 12035 (1996);
 A. Chakrabarti and B. Bhattacharyya, {\it ibid}. {\bf 56}, 13786
 (1997);
 Y. Liu, Z. Hou, P. M. Hui, and W. Sritrakool, {\it ibid}. {\bf 60},
 13444 (1999).

%--- random wire experiment
\bibitem{13}
 R. G. Steinmann and B. Pannetier, Europhys. Lett. {\bf 5}, 559 (1988).

%--- rhombus E-spec theory
\bibitem{14}
 J. Vidal, R. Mosseri, and B. Dou\c{c}ot, Phys. Rev. Lett. {\bf 81},
 5888 (1998).

%--- rhombus sc wire experiment
\bibitem{15}
 C. C. Abilio, R. Butaud, Th. Fournier, B. Panneiter, J. Vidal, S.
 Tedesco, and B. Dalzotto, Phys. Rev. Lett. {\bf 83}, 5102 (1999).

%--- rhombus E-spec without magnetic field
\bibitem{16}
 B. Sutherland, Phys. Rev. B {\bf 34}, 5208 (1986).

\end{references}
\end{document}